\documentclass[10pt,twocolumn]{article}
\usepackage{amsmath,amssymb}
\usepackage{citesort,epsfig}
\begin{document}

\title{Distribution of the second virial coefficients of globular
proteins}
\author{{\bf Richard P. Sear}\\
Department of Physics, University of Surrey,
Guildford, Surrey GU2 7XH, United Kingdom\\
{\tt r.sear@surrey.ac.uk}}
\maketitle

\begin{abstract}
George and Wilson [Acta. Cryst. D 50, 361 (1994)] looked at the
distribution of values of the second virial coefficient
of globular proteins, under the conditions at which they crystallise.
They found the values to lie within a fairly narrow range.
We have defined a simple model of a generic globular protein.
We then generate a set of proteins by picking values for the
parameters of the model from a probability distribution. At fixed
solubility, this set of proteins is found to have values
of the second virial coefficient that fall within a fairly narrow
range. The shape of the probability distribution
of the second virial coefficient is Gaussian because the
second virial coefficient is a sum of contributions from different
patches on the protein surface.
\end{abstract}


Protein crystallisation is an important problem yet our grasp of the
details of how it occurs is very poor. Proteins need to be crystallised
from solution in order to
determine their structure
via X-ray crystallography \cite{durbin96,piazza00}.
The crystallisation presumably starts with heterogeneous nucleation
of the crystalline phase in the protein solution, but there has been no
systematic experimental study of this, as far as the author is aware.
Without an understanding of how proteins crystallise, protein crystallisation
is almost totally {\em ad hoc}: essentially the only way to know
if a protein will crystallise under a certain set of conditions is to
try it. It would be enormously useful if we could {\em predict}
the conditions
under which a protein was most likely to crystallise.
Here by protein we mean globular protein, which are proteins
that are soluble in solution, as opposed to membrane proteins
which exist embedded in a surfactant bilayer.
The hope
that it is possible to predict the conditions that promote crystallisation
motivated George and Wilson \cite{george94} to look at
the values of the (osmotic) second virial coefficient of a number of
proteins under the conditions where they were crystallised.
They found that the second virial coefficient was always negative
and lay within what they called `a fairly narrow range'.
If we ignore outliers then second virial coefficients
gathered together by Haas and Drenth \cite{haas98},
and converted to reduced units by Vliegenthart and
Lekkerkerker \cite{vliegenthart00},
lie in the range $-8$ to
$-40$, in units of the volume of the protein; see Table V of
Ref.~\cite{vliegenthart00}.

The simplest
explanation of this range is that the upper limit
is set by the requirement that the attractive interactions be strong
enough to pull the molecules into a crystal from a dilute solution.
The lower limit is set by the dynamics of the solution, if the
attractive interactions are too strong the protein molecules
tend to aggregate irreversibly and this aggregation preempts and
prevents crystallisation. Testing these explanations is all but
impossible due to our poor understanding of crystallisation
so we turn to a well-defined, and easily calculated,
property of a protein solution: its solubility.
We consider the solubility
of the protein, i.e., the concentration of protein in the solution
which coexists with the crystal, in preference to the process
of crystallisation.
We ask the question: For a given solubility, say 5\% by
volume, what is the distribution of values that we expect for $B_2$?
If we have 1000 proteins, say, all with the same solubility,
then is their distribution of values of $B_2$ very broad, or is it
narrow? What is the shape of the distribution, i.e., what is
its functional form?

The distribution of values of $B_2$ of a large number of proteins
defines a probability distribution function $P(B_2)$. We
will consider a constraint, that of fixed solubility, and so
will obtain a probability distribution function that also
depends on this constraint.
We are
inspired to study this function by a range of work on protein
solutions and crystals
\cite{sear99,neal99,haas99,curtis01,carlsson01,kulkarni,warren}
that has shown that protein-protein
interactions are well described by a sum
over contacts between the proteins. Where by a contact 
between two proteins,
we mean that
a specific patch on the surface of one protein approaches closely,
a couple of \AA, to a specific patch on the surface of the other protein.
Now, if these contacts are more-or-less independent then we expect them
to contribute essentially independently to $B_2$. But if $B_2$
is the sum of many independent contributions for each protein, then
we know the form of the distribution function $P(B_2)$: it
is a Gaussian. The central limit theorem states that the probability
distribution function of some property $Y$, which is a
sum over a large number of independent random variables, is a Gaussian
\cite{ma}. Also of course the more independent patches there are
on the surface the narrower will be the distribution of values of $B_2$.

The physical picture is that the surface of a protein
has a number of patches on its surface.
Under the conditions where the protein's
solubility is low, these patches attract each other.
The strength of each patch
attraction is then a random variable selected from some
distribution. It is a random variable
if the strength of the attraction of one patch on the surface
is independent of the attraction of any of the other
patches. This physical picture is very simple
and is of course approximate but if the correlations between the
various patches are weak then $P(B_2)$
should be approximately Gaussian. Of course this will also apply
to the probability distribution function of any other variable which is
a sum of more-or-less independent contributions from the surface patches.
We do not know whether the rate of crystallisation is such a variable.

Before we consider our model it is worthwhile noting that within biology
there is a move away from studying proteins one at a time to
studying them en masse, e.g., studying the complete proteome of an
organism. Where the proteome is defined as being the complete set of
proteins possessed by an organism.
This follows on from work on establishing the complete genome of a number
of organisms \cite{hugo,apweiler01}. Although here we consider sets
of proteins which are just those we want to crystallise, and so
can come from a number of different organisms, in the future
the solubilities and virial
coefficients of complete proteomes could be considered.

\begin{figure}[t]
\begin{center}
\caption{
\lineskip 2pt
\lineskiplimit 2pt
A schematic of our model protein. It is drawn as a cube with
the attractive patches drawn as black patches on the faces of
the cube. The model occupies 2 by 2 by 2 $=8$ lattice sites. The
individual lattice sites occupied by the model are separated by the
thinner lines.
\label{model}
}
\vspace*{0.1in}
\epsfig{file=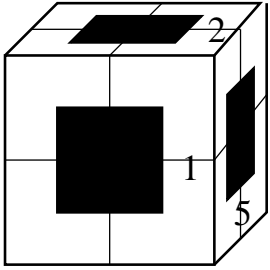,width=1.4in}
\end{center}
\end{figure}

The model is chosen to be as simple as possible,
while incorporating the patchy nature of the surface of proteins
together with the variability in the interactions from protein to protein.
Thus for simplicity we chose a lattice model. The lattice
is cubic and each protein occupies 8 lattice sites arranged
2 by 2 by 2,
see Fig.~\ref{model}. We make the model 2 sites across to reduce the
range of the attraction, which is 1 site, to half the diameter
of the hard core.
The model `proteins' can rotate, and so
have 24 distinct orientations. Each of the 6 faces of the cube
has a patch, labeled $i=1$ to 6, with patches 1 to 4 clockwise around
a loop of 4 of the faces, and patches 5 and 6 on the remaining
2 faces.
The interactions between
model proteins are pairwise additive and consist of 2 parts.
The first is simply an excluded volume interaction:
2 proteins cannot overlap.
The second is that if the faces of 2 proteins
are in contact
there is an energy of interaction between the 2 touching patches of the
2 proteins. By in contact we mean that {\em all} of the 4 lattice sites on the
face of one protein are in contact with one of the lattice sites
of the face of the other protein, in other words the faces must
overlap completely otherwise the energy of interaction is taken to be zero.
The touching patches are those on the faces of the 2 proteins
that face each other.
If the $i$th patch of one protein is adjacent to the $j$th
patch of another protein then there is an interaction energy of
$\epsilon_{ij}$.
Different proteins will have the same excluded
volume interactions but the set of interactions $\epsilon_{ij}$ will be
different to represent the different surfaces of different proteins.
The interactions form a symmetric square matrix,
$\epsilon_{ij}=\epsilon_{ji}$.

Thus, a protein is specified by giving values to the 21 distinct
$\epsilon_{ij}$; these are composed of 36 interactions,
consisting of 6 like interactions, $i=j$, and 30
unlike interactions, $i \ne j$, with 15 of these related to the
other 15 by symmetry. The energies $\epsilon_{ij}$ determine the
phase behaviour via Boltzmann weights $b_{ij}=\exp(-\epsilon_{ij}/kT)$,
where $k$ is Boltzmann's constant and $T$ is the temperature. We will
in fact deal mainly with these weights not with the energies themselves.

For the values of the $b_{ij}$ for the patch-patch interactions
we again choose the simplest possible model and neglect any correlations
Each of the 21 distinct
$b_{ij}$ is {\em almost} a stochastic variable described by a probability
function $p(b)$, i.e., the probability that $b_{ij}$
lies between $b$ and $b+{\rm d}b$ equals
$p(b){\rm d}b$. They are almost but not quite uncorrelated because
we want the crystalline phase of all our 
model proteins to be a simple cubic lattice with 1 protein molecule
per unit cell. The crystal is close-packed; there is a protein
on every lattice site.
In such a crystal, 3 different bonds are formed:
the $13$, $24$ and $56$ bonds and the energy $e_x$ is
\begin{eqnarray}
e_x&=&\epsilon_{13}+\epsilon_{24}+\epsilon_{56}\nonumber\\
\frac{e_x}{kT}&=&-\ln\left(b_{13}b_{24}b_{56}\right)
\label{ex}
\end{eqnarray}
We want this to be the ground state and so having generated the
21 distinct Boltzmann weights, we find the 3 largest of these,
call them $b_{\alpha\beta}$, $b_{\gamma\delta}$ and
$b_{\delta\zeta}$ and then we interchange
$b_{13}$ and $b_{\alpha\beta}$, $b_{24}$ and $b_{\gamma\delta}$
and $b_{56}$ and $b_{\delta\zeta}$. In this way we ensure
the simple cubic lattice is the ground state of the model
protein. Performing this swapping procedure introduces correlations of
course, but they are minor as we have ordered only 3 of the
21 weights.

The entropy of the crystalline
phase is zero, as the cubes can neither translate nor rotate, so its
free energy is equal to its energy, $e_x$.
At low temperature, the pressure will be low and so the chemical
potential will be closely equal to the free energy, which we have
already said is equal to the energy. Thus the chemical potential
in the crystalline phase $\mu_x\approx e_x$. The chemical potential
in a dilute solution, which is well described by an ideal gas, is
\begin{equation}
\mu_{ig}=kT\ln(\rho/24),
\label{muv}
\end{equation}
where the 24 comes from the rotational entropy $k\ln 24$ of a freely
rotating cube with all 6 faces distinct. $\rho$ is the number density,
the number of proteins per lattice site.
When a dilute solution coexists with the
crystal then we can treat the solution as an ideal gas and find the
density of this solution by equating $\mu_{ig}$ of Eq.~(\ref{muv}) with
$\mu_x$, which is closely equal to $e_x$ of Eq.~(\ref{ex}). So, the
density of the solution that coexists with the crystal, i.e., the
solubility of the protein is
\begin{equation}
\rho_s=24\exp(e_x/kT).
\label{rhosol}
\end{equation}
A more useful measure of density is the volume fraction $\phi=8\rho$
which the is the fraction of the sites in the fluid occupied by the
proteins. Thus we will work in terms of the solubility volume
fraction $\phi_s=8\rho_s$.

The second virial coefficient of a continuum model is defined
by
\begin{equation}
B_2=-\frac{1}{2}\int
{\rm d}{\bf r}_{12}{\rm d}{\bf\omega}_1{\rm d}{\bf\omega}_2
\left[\exp\left(-u({\bf r}_{12},{\bf\omega}_1,{\bf\omega}_2)\right)-1\right],
\end{equation}
where ${\bf r}_{12}$ is the distance between the centres of mass
of two molecules, and ${\bf \omega}_1$ and ${\bf\omega}_2$
are the orientations of molecules 1 and 2, respectively.
The integrations over the angles are normalised.
$u({\bf r}_{12},{\bf \omega}_1,{\bf\omega}_2)$ is the energy
of interaction of pair of molecules, as a function of their
separation and orientation. For a lattice model the integrals
are replaced by sums and we have
\begin{equation}
B_2=-\frac{1}{1152}\sum_{\alpha=1}^{24}\sum_{\beta=1}^{24}
\sum_{\left\{ {\bf r}_{12} \right\}}
\left[\exp\left(-u({\bf r}_{12},\alpha,\beta)\right)-1\right],
\label{bdef1}
\end{equation}
where the three sums are, in order, over the 24 orientations of
molecule 1, over the 24 orientations of molecule 2 and over the
separation of the two molecules. The factor of 1152 comes from the
factor of half and two factors of 24 from the normalisation of the
sums over orientation.

\begin{figure}[t]
\begin{center}
\caption{
\lineskip 2pt
\lineskiplimit 2pt
A scatter plot of the
the solubility volume fraction $\phi_s$, the $y$ axis,
against the reduced second virial coefficient $B_2/B_{2hc}$, the $x$ axis,
for 10,000
proteins. The mean weight
${\overline b}=9$ and the distribution is a top-hat with
width $b_w=16$.
\label{w2}
}
\vspace*{0.3in}
\epsfig{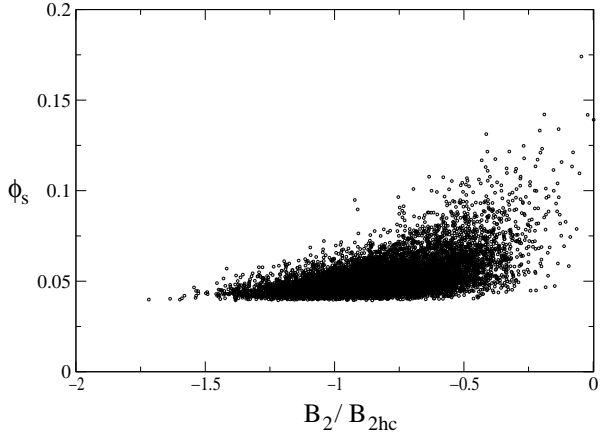}
\end{center}
\end{figure}

The second virial coefficient $B_2$ of the protein-protein interaction
can be calculated once the set of weights $b_{ij}$ are specified.
The exponential factor in Eq.~\ref{bdef1} equals 0 if the two
molecules overlap and equals one of the $b_{ij}$
when the molecules occupy adjacent sites. Thus we have
\begin{equation}
B_2=\frac{1}{2}\left[27-\frac{1}{6}
\sum_{i=1}^6\sum_{j=1}^6\left(b_{ij}-1\right)\right],
\label{b2def}
\end{equation}
where the first term inside the brackets comes from excluded volume
interactions and the second from the interactions between touching patches.
The number 27 comes from the fact that each model protein excludes
other proteins from a cube of 3 by 3 by 3 lattice sites. Thus,
in the high temperature limit where $b_{ij}=1~~\forall i,j$,
$B_2=B_{2hc}=27/2$.
The sums over 24 orientations reduce to sums over 6 orientations
as rotating either of the 2 molecules around the axis joining
their centres does not change the energy. The factor in front
of the double sum is a normalisation factor of $1/36$ times the
6 possible lattice sites that one molecule can occupy and be
adjacent to the other molecule.

Finally, for the purposes of performing example calculations we will take the
distribution function $p(b)$ to be a top-hat function of mean
$\overline{b}$ and width $b_w$.
\begin{equation}
p(b)=\left\{
\begin{array}{cc}
0 & b < \overline{b}-b_w/2 \\
b_w^{-1} &\overline{b}-b_w/2 < b
< \overline{b}+b_w/2 \\
0 & b > \overline{b}+b_w/2
\end{array}\right.   .
\label{distf}
\end{equation}

With our model defined we can generate a protein by generating values
for the 21 distinct Boltzmann weights for the
patch-patch interactions, $b_{ij}$, according
to the probability distribution function $p(b)$. Repeating
this procedure many times will generate a set of proteins, each
protein having
a distinct array of interactions, $b_{ij}$.
This set can then be analysed
to look, for example, for correlations between different properties.

Before we look at a set of proteins with solubilities in a fixed, small,
range let us look at just a set of proteins, with a range of solubilities.
Fixing the mean patch-patch interaction weight ${\overline b}=9$
and the width $b_w=16$ so that the Boltzmann weights lie in the
range $1$ to 17, we have generated a set of 10,000 proteins.
Humans have about 10 times as many different proteins as this,
bacteria typically less than half this number \cite{apweiler01}.
The mean weight was chosen to give
a distribution centred around $\phi_s\approx 0.05$. Of course
increasing ${\overline b}$ will shift the distribution to lower values
of $\phi_s$ and decreasing ${\overline b}$ will shift the
distribution to higher values of $\phi_s$.
Their solubilities and second virial coefficients are plotted as
a scatter plot in Fig.~\ref{w2}. Clearly, there is a correlation between
the second virial coefficient of a protein and its solubility: the
more negative is $B_2$ the lower is the solubility on average, but
there is also considerable scatter.

\begin{figure}[t]
\begin{center}
\caption{
\lineskip 2pt
\lineskiplimit 2pt
A plot of the probability distribution function $P(B_2/B_{2hc})$ of the
reduced second virial coefficient $B_2/B_{2hc}$,
for model proteins with a solubility
in the range $\phi_s=0.05\pm0.01$. As in Fig.~\ref{w2},
${\overline b}=9$ and $b_w=16$.
\label{b2h}
}
\vspace*{0.3in}
\epsfig{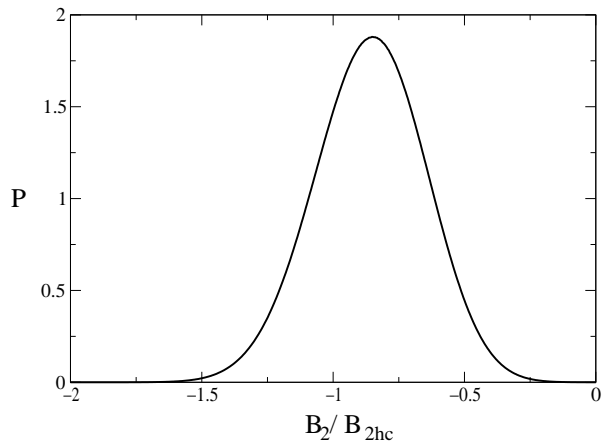}
\end{center}
\end{figure}

Now, restricting the solubility of a protein to lie in the narrow range
$\phi_s=0.05\pm0.01$, we have calculated the probability distribution of
second virial coefficients and plotted the result in Fig.~\ref{b2h}.
The curve is almost Gaussian; a Gaussian is a good fit to the data.
The standard deviation is close to $0.2$.
We have used
a top-hat probability distribution function for the Boltzmann
weights of patch-patch interactions, Eq.~\ref{distf}, but
of course our finding of a Gaussian distribution is insensitive
to the exact form of this distribution function.

There is relatively little data for proteins where both the
second virial coefficient and the solubility have been measured.
Extensive results are available for lysozyme
\cite{guo99,rosenbaum99}, but there are results
for both the virial coefficient and the solubility for only
a couple of other proteins. Rosenbaum {\em et al.}
\cite{rosenbaum99} plot the reduced second virial coefficient
against solubility of lysozyme under a number of different
conditions, $\gamma$-crystallin and
BPTI (bovine pancreatic trypsin inhibitor), their Fig. 4.
With only 3 proteins no attempt can be made at assessing
a probability distribution. However, in the wake of the sequencing
of whole genomes there is a drive towards so-called `high throughput'
methods which can rapidly assess the properties of
large numbers of proteins. If such a
method could be developed for second virial coefficients the results
could be compared to our predictions. 
For a given protein there is clear experimental evidence
that as conditions are varied so as the make the second virial
coefficient more negative the solubility decreases \cite{poon00,kulkarni}.
For our model, this corresponds to increasing the mean Boltzmann weight
of the attractions, ${\overline b}$, which will decrease both $B_2$ and
the solubility, whatever the distribution of weights for the patch-patch
attractions.

We have swapped 3 of the patch-patch interaction energies
to force the ground state to be a simple cubic lattice.
Because of this, when the attractions
are made strong enough, ${\overline b}$ sufficiently
positive, our model proteins all have low solubilities. Thus they
are presumably representative
of proteins that are easily crystallisable. If we had not
swapped the 3 interaction energies, and we assume that
the only possible crystalline phase is the simple cubic, then
some of our model proteins will be highly soluble: they will
not crystallise from dilute solutions.

In conclusion,
the surfaces of proteins are patchy and mediate short-range interactions,
short with respect to the size of the protein, which is a few nms.
A quantity such as the second virial coefficient is an integral (sum
for lattice models) over the contribution of the core of the protein,
over the contributions of
any attractions between patches of the surface, and
over the contribution of any
longer-ranged interactions such as an overall
electrostatic repulsion.
If the patches are independent
or almost independent, then the distribution function for their total contribution
to the second virial coefficient, of
a large number of proteins of a similar size will tend to be
Gaussian. A straightforward consequence of the central limit
theorem.
The central limit theorem applies
to a variable, here the
second virial coefficient, that is the sum of a large number of
random variables
with the same mean and variance.  The suggestion is that if experiments
are done on a large set of proteins all at the same solubility
and all with similar sizes, then
this fixed solubility will set a rough scale for the mean strength
of the interactions. Then if proteins are highly modular, the
predominant variation from protein to protein will come from
variations in the sum of the patch-patch Boltzmann weights, and
so will have a roughly Gaussian distribution. With a width roughly
equal to the the square root of the number of independent patches
times the width in the distribution of the Boltzmann weight of a single
patch.
Thus, a second
virial coefficient which is a sum over contributions from a number
of independent patches will tend to have a narrow distribution of values.
This may be partly responsible for the fairly narrow range
of values of the second virial coefficient over which most proteins
crystallise \cite{george94,haas98,vliegenthart00}
but crystallisation is both complex and
poorly understand so there are almost certainly other factors.
By highly modular we mean that the patches on a protein's
surface are close to being independent. If they were truly independent
we would have
$<b_{ij}b_{i'j'}>={\overline b}^2$, $ij\ne i'j'$,
where $<>$ indicates an average over all the proteins, and the
interactions $ij$ and $i'j'$ are any two patch-patch
interactions of the same protein. We are assuming that
violations of this equality are weak.

It is a pleasure to acknowledge discussions with
J. Cuesta, D. Frenkel and P. Warren.

\end{document}